\documentclass[aps,prd,10pt,twocolumn,amsmath,amssymb]{revtex4-1}
\pdfoutput=1

\usepackage{natbib}
\usepackage{graphicx}
\usepackage{dcolumn}
\usepackage{bm}

\begin{document}


\let\phi\varphi
\def\d{\mathrm{d}}
\def\p{\partial}
\def\be{\begin{equation}}
\def\ee{\end{equation}}
\def\bea{\begin{eqnarray}}
\def\eea{\end{eqnarray}}
\def\SdS{Schwarz\-schild--de~Sitter }
\def\RNdS{Reissner--Nordstr\"{o}m--de Sitter }
\def\KdS{Kerr--de~Sitter }
\def\KNdS{Kerr--Newman--de~Sitter }


\title{Test particle motion along equatorial circular orbits in the revisited \KdS spacetime}


\author{Petr Slan\'{y}}
	\email{petr.slany@physics.slu.cz}
\affiliation{Research Centre for Theoretical Physics and Astrophysics, Institute of Physics, 
Silesian University in Opava\\ Bezru\v{c}ovo n\'{a}m. 13, CZ-746\,01 Opava, Czech Republic}%


\begin{abstract}
Both circular and epicyclic motion of test particles along equatorial circular orbits in the revisited \KdS spacetime is analyzed. We present relations for specific energy, specific angular momentum and Keplerian angular velocity of particles on equatorial circular orbits, and discuss criteria for the existence and stability of such orbits giving limits on spacetime parameters. Finally, we discuss the epicyclic motion along equatorial circular orbits obtaining relations for radial and vertical epicyclic frequencies. The results are compared with those for standard \KdS geometry.
\end{abstract}


\maketitle


\section{Introduction}
There is a strong observational evidence that our Universe expands and during the last four billions years its expansion is accelerated \cite{Rie-etal:1998:ASTRJ1:,Per-etal:1999:ASTRJ2:,Jon-etal:2019:ASTRJ2:,Planck_col:2016:ASTRA:,Boss_col:2015:PHYSR4:}. The real cause of this phenomenon is still unknown and was named \emph{dark energy}. In terms of relativistic cosmology, the dark energy is well described by a positive cosmological constant $\Lambda$, characterizing any fluid with equation of state $p=-\epsilon$, where $p$ and $\epsilon$ denote pressure and energy density of the fluid, respectively. Moreover, the same equation of state is valid also for Lorentz invariant vacuum energy density \cite{Gro-Her:2007:GRwCosmology:}. Therefore, it seems natural (and being closer to reality) to study processes around black holes in the asymptotically de~Sitter rather than flat backgrounds. 

Concentrating on rotating uncharged black holes, the most famous is the \KdS (KdS) solution of vacuum Einstein field equations with cosmological term described by B. Carter \cite{Car:1973:BlaHol:}. This solution is characterized by constant curvature scalar in the whole spacetime, $R=4\Lambda=\mbox{const}$. Recently, a new exact solution of Einstein field equations describing rotating (Kerr) black hole in de~Sitter universe was found \cite{Ova-Con-Stu:2021:PHYSR4:}. The authors used the method of so-called gravitational decoupling \cite{Ova:2017:PHYSR4:,Ova:2019:PHYLB:,Con-Ova-Cas:2021:PHYSR4:} applied to special form of the Kerr-Schild metric, namely the G\"{u}rses-G\"{u}rsey metric \cite{Gur-Gur:1975:JMATP4:}, to obtain a rotating, axially-symmetric version of the \SdS metric. This \emph{revisited} \KdS metric is simpler in its form but the spacetime structure seems to be richer in comparison with the standard KdS geometry. As Ovalle et al. showed, the main difference is in non-uniformity of curvature scalar and its dependence on black-hole spin, being axially symmetric as well as plane-symmetric with respect to the equatorial plane; in the equatorial plane, then, the curvature scalar has the same constant value as in the KdS spacetime. Since this ``warped effect'' is significant only near rotating black holes, it could demonstrate some influence of strong gravity (and, particularly, of frame dragging) on vacuum energy \cite{Ova-Con-Stu:2021:PHYSR4:,Ova:2022:EPJC:}. On the other hand, in contrast to the standard KdS metric, the revisited KdS form does not contain terms proportional to $\Lambda a^2$, where $a=J/cM$ is the specific angular momentum (spin) of the black hole, which could be interpreted, contrarily, as a~weakening of coupling between rotation and cosmic repulsion. In general, the revisited \KdS metric can describe the spacetime of rotating black hole immersed in dark-energy bath (of any kind) with anisotropic equation of state, characterized by the stress-energy tensor satisfying the dominant energy condition, see \cite{Ova-Con-Stu:2021:PHYSR4:,Ova:2022:EPJC:} for more details. 

In this paper, we analyze test particle motion along equatorial circular orbits in the revisited KdS spacetime of both black holes and naked singularities. Moreover, the epicyclic motion, following from small linear perturbations of particles on equatorial circular orbits, is discussed. Equatorial circular orbits (geodesics) of massive particles and photons in standard KdS spacetime were intensively studied in past decades, see, e.g., \cite{Stu-Sla:2004:PHYSR4:,Kra:2004:CLAQG:,Kra:2005:CLAQG:,Hak-etal:2010:PHYSR4:,Char-Stu:2017:EURPJC:,Stu-Char-Sche:2018:EURPJC:}. Astrophysical processes influenced by the relict cosmological constant are discussed in \cite{Stu:2005:MODPLA:,Stu-Sche:2011:JCAP:,Stu-etal:2020:UNIVERSE:}. Therefore, we also compare obtained results with those for the standard KdS spacetime. Note that differences in black-hole shadows for standard and revisited KdS backgrounds were analyzed and discussed in \cite{Ova-Con-Stu:2021:PHYSR4:,Omw-etal:2022:EPJC:}.

\section{Revisited \KdS spacetime}
In the standard Boyer-Lindquist coordinates $(t,\,r,\,\theta,\,\phi)$ and geometrical units ($c=G=1$), the revisited \KdS spacetime (rKdS) is described by the line element \cite{Ova-Con-Stu:2021:PHYSR4:}
\bea\label{e1}
\lefteqn{\d s^2=-\frac{\Delta}{\rho^2}(\d t-a\sin^2 \theta\d\phi)^2 + } \\
& & \frac{\sin^2 \theta}{\rho^2}\left[a\,\d t- \left(r^2+a^2 \right) \d\phi \right]^2 + \frac{\rho^2}{\Delta} \d r^2 + \rho^2\d\theta^2, \nonumber
\eea
where
\bea
     \Delta & = & r^2 -2Mr + a^2 -\frac{\Lambda}{3}r^4,   \label{e2} \\
     \rho^2 & = & r^2 +a^2 \cos^2 \theta.                 \label{e3}
\eea
Given spacetime is characterized by three parameters: $M$ and $a>0$ correspond to the mass and spin of the center, $\Lambda>0$ is a cosmological constant. Further it is convenient to introduce instead of $\Lambda$ a dimensionless cosmological parameter 
\be\label{e4}
y=\frac{\Lambda}{3}M^2
\ee
and put $M=1$ in order to get all the quantities dimensionless.

\begin{figure}
\includegraphics[width=1 \hsize]{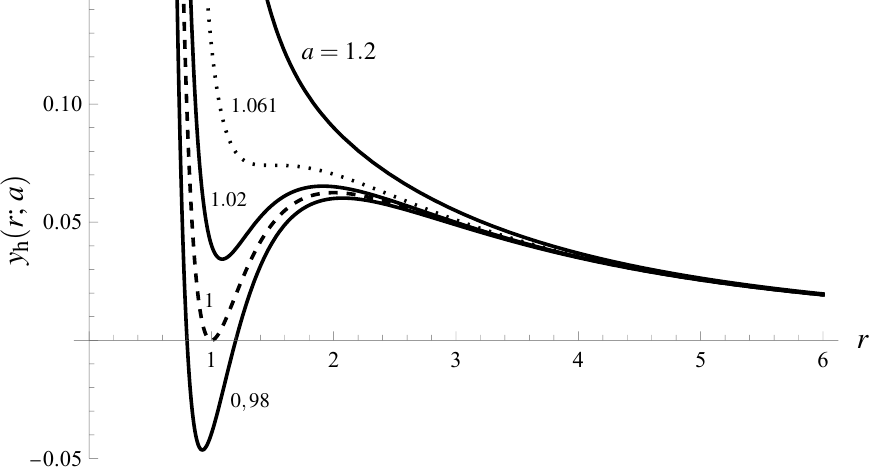}
\caption{Profiles of the function $y_{\rm h}(r;\,a)$ for various values of the parameter $a$. Dashed profile corresponds to the case of $a=1$. Dotted profile corresponds to the case of $a=a_{\rm c(bh)}\doteq 1.061$, for which both local extrema coincide in an inflection point determining maximal value $y_{\rm c(bh)}\doteq 0.074$ of the cosmological parameter $y$ enabling existence of rKdS black holes.}
\label{f1}
\end{figure}

Horizons of the spacetime correspond to the null hypersurfaces of $r=\mbox{const.}$ given by the condition $g^{rr}=0$, i.e. $\Delta=0$, which can be written in the form
\be\label{e5}
y=y_{\rm h}(r;\,a)\equiv\frac{r^2-2r+a^2}{r^4}.
\ee
Local extrema of the function $y_{\rm h}(r;\,a)$ are given by the condition
\be\label{e5.1}
a^2=a^2_{\rm ex(h)}(r)\equiv \frac{1}{2}r(3-r).
\ee
Analysis of the function $a^2_{\rm ex(h)}(r)$ shows that for $0<a<a_{\rm c(bh)}=3\sqrt{2}/4\doteq 1.061$, the function $y_{\rm h}(r;\,a)$ has two local extrema: minimum $y_{\rm h,min}$ (which for $a<1$ is negative) and maximum $y_{\rm h,max}$. Thus, for $0<y<y_{\rm h,max}$ (if $a<1$) or $y_{\rm h,min}<y<y_{\rm h,max}$ (if $1<a<a_{\rm c(bh)}$) the spacetime contains three horizons in the region $r>0$: the inner and outer black-hole horizons, $r_{-}$ and $r_{+}$, and the cosmological horizon $r_{\rm c}$, where $r_{-}<r_{+}<r_{\rm c}$. For $y=y_{\rm h,min}$ (which for $1<a<a_{\rm c(bh)}$ is positive), both black-hole horizons coincide, $r_{-}=r_{+}<r_{\rm c}$; such situation corresponds to an extreme rKdS black hole. All other geometries correspond to naked ring singularities lying in the equatorial plane at $r=0$. For $a=a_{\rm c(bh)}$, both extrema of the function $y_{\rm h}(r;\,a)$ coincide in the inflection point located at $r=3/2$. Corresponding value of the cosmological parameter $y$ determines its maximal value $y_{\rm c(bh)}=2/27$ allowing existence of black holes in the rKdS spacetimes. Radial profiles of the function $y_{\rm h}(r;\,a)$ for various values of the rotational parameter $a$ are shown in the Fig.~\ref{f1}. Note that behavior of the function $y_{\rm h}(r;\,a)$ and characteristic critical values of spacetime parameters allowing black holes are the same as in the case of \RNdS spacetime, where the charge parameter $e$ instead of the rotational parameter $a$ influences behavior of the function of horizons $y_{\rm h}(r)$ \cite{Stu-Hle:2002:ACTPS2:}. On the other hand, in the standard KdS spacetime, the critical values of spacetime parameters enabling existence of black holes are $y_{\rm c(bh)}^{\rm KdS}=16/(3+2\sqrt{3})^3$ and $a_{\rm c(bh)}^{\rm KdS}=\sqrt{3(3+2\sqrt{3})}/4$ \cite{Stu-Sla:2004:PHYSR4:}, being slightly different from those in the rKdS spacetime.

\section{Equatorial circular orbits}
The metric of rKdS spacetime (\ref{e1}) has the same form as the metric of Kerr or Kerr--Newman spacetimes \cite{Mis-Tho-Whe:1973:Gra:}. The only difference is in the expression of $\Delta$. All these spacetimes are stationary and axially symmetric, which means that they possess two Killing vector fields, $\xi^{\mu}=\delta^{\mu}_{t}$ and $\eta^{\mu}=\delta^{\mu}_{\phi}$. Moreover, corresponding Hamilton--Jacobi equations describing motion of test particles in given backgrounds can be written in a~separated form. This crucial property was found by B.~Carter \cite{Car:1968:PHYSREV:,Car:1973:BlaHol:}. For uncharged test particles, Carter's equations reads:\footnote{Their generalization for the motion of charged test particles in the \KNdS spacetime can be found in \cite{Stu:1983:BULAI:,Sla-Stu:2020:EPJC:}.}
\bea
\rho^2 \frac{\d r}{\d\lambda} &=& \pm\sqrt{R(r)}, \label{e6} \\
\rho^2 \frac{\d\theta}{\d\lambda} &=& \pm\sqrt{W(\theta)}, \label{e7} \\
\rho^2 \frac{\d t}{\d\lambda} &=& \frac{(r^2+a^2)}{\Delta}P_r - aP_{\theta}, \label{e8} \\
\rho^2 \frac{\d\phi}{\d\lambda} &=& \frac{a}{\Delta}P_r -\frac{P_{\theta}}{\sin^2\theta}, \label{e9}
\eea
where
\bea
R(r) &=& P_r^2 - \Delta(m^2r^2+\mathcal{K}), \label{e10} \\
P_r &=& (r^2+a^2)\mathcal{E}-a\varPhi, \label{e11} \\
W(\theta) &=& \mathcal{K}-m^2a^2\cos^2\theta - \frac{P_{\theta}^2}{\sin^2\theta}, \label{e12} \\
P_{\theta} &=& a\mathcal{E}\sin^2\theta -\varPhi.  \label{e13}
\eea
The motion is characterized by four constants of motion given by corresponding projections of particle's 4-momentum $p^{\mu}=\d x^{\mu}/\d\lambda$: ``energy'' $\mathcal{E}=-p^{\mu}\xi_{\mu}$, ``angular momentum'' $\varPhi=p^{\mu}\eta_{\mu}$, rest mass-energy $m=(-p^{\mu}p_{\mu})^{1/2}$, and Carter's constant $\mathcal{K}=\xi_{\mu\nu}p^{\mu}p^{\nu}$ (connected with the existence of certain Killing tensor $\xi_{\mu\nu}$ \cite{Wal:1984:GenRel:}).

\begin{figure*}
\centering
\includegraphics[width=.8 \hsize]{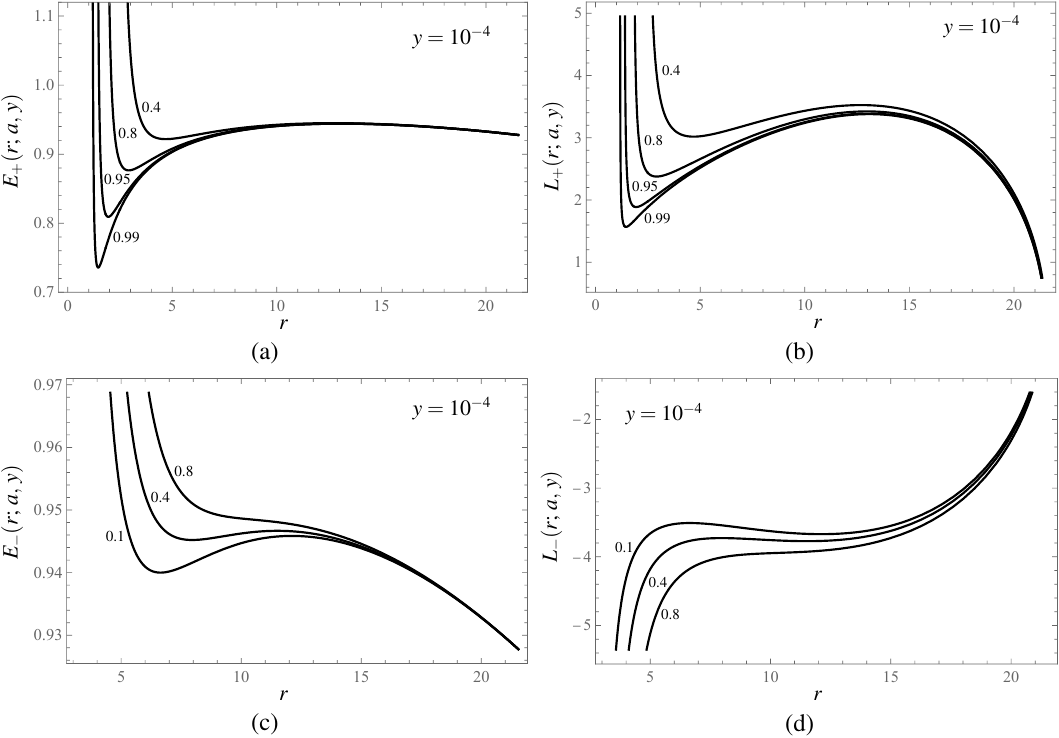}
\caption{Radial profiles of specific energy and specific angular momentum of test particles moving on equatorial circular orbits around rKdS black holes with cosmological parameter $y=10^{-4}$. Various profiles correspond to various values of the rotational parameter $a$ assigned to each curve. Panels (a) and (b) correspond to the plus-family orbits, panels (c) and (d) correspond to the minus-family orbits.}
\label{f2}
\end{figure*}

Further we will concentrate on test-particle motion in the equatorial plane $\theta=\pi/2$. Such a motion is characterized by the constant $\mathcal{K}=(a\mathcal{E}-\varPhi)^2$. Moreover, instead of the affine parameter $\lambda$ we use particle's proper time $\tau=m\lambda$ and define ``specific energy'' and ``specific angular momentum'' of a particle by relations: $E=\mathcal{E}/m$, $L=\varPhi/m$. The equation of radial motion (\ref{e5}), therefore, takes the form
\be\label{e14}
r^2 \frac{\d r}{\d\tau} = \pm\sqrt{\tilde{R}(r)},
\ee
where
\be\label{e15}
\tilde{R}(r)=[(r^2+a^2)E-aL]^2-\Delta[r^2+(aE-L)^2].
\ee

\begin{figure*}
\centering
\includegraphics[width=.8 \hsize]{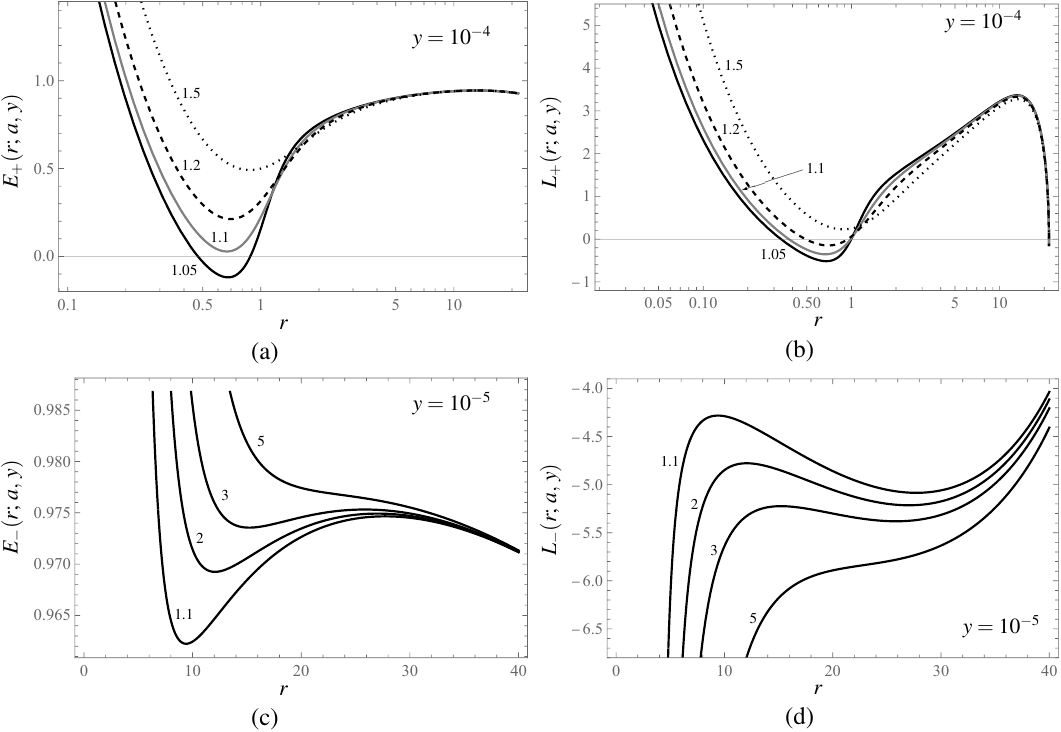}
\caption{Radial profiles of specific energy and specific angular momentum of test particles moving on equatorial circular orbits around rKdS naked singularities. Various profiles correspond to various values of the rotational parameter $a$. Panels (a) and (b) correspond to the plus-family orbits, panels (c) and (d) correspond to the minus-family orbits.}
\label{f3}
\end{figure*}

Circular orbits $r=\mbox{const}$ are given by conditions $\tilde{R}=0$ and $\d\tilde{R}/\d r=0$, which must hold simultaneously. Analysis of these conditions, performed in the same way as in the standard KdS spacetime \cite{Stu-Sla:2004:PHYSR4:}, enables to obtain relations for specific energy and specific angular momentum of particles moving on equatorial circular orbits in revisited \KdS spacetime:
\be\label{e16}
E_{\pm}(r;\,a,\,y) = \frac{\displaystyle 1-\frac{2}{r}-yr^2\pm a\sqrt{\frac{1}{r^3}-y}}{\displaystyle\left[ 1-\frac{3}{r}\pm 2a\sqrt{\frac{1}{r^3}-y}\right]^{1/2}},
\ee
\be\label{e17}
L_{\pm}(r;\,a,\,y) = \frac{\displaystyle -\frac{2a}{r}-yar^2\pm (r^2+a^2)\sqrt{\frac{1}{r^3}-y}}{\displaystyle\left[ 1-\frac{3}{r}\pm 2a\sqrt{\frac{1}{r^3}-y}\right]^{1/2}}.
\ee
Note that these relations are valid also for $y<0$, i.e. for the attractive cosmological constant $\Lambda<0$. Characteristic behavior of specific energy and specific angular momentum for particles on circular orbits in the equatorial plane around rKdS black holes and naked singularities is shown in Figs. \ref{f2}, \ref{f3}. Due to dragging of frames caused by non-zero source spin $a>0$, we can distinguish two families of orbits, the plus-family and the minus-family (corresponding to the upper and lower signs in relations (\ref{e16}) and (\ref{e17})).
We will discuss their orientation and stability later. First we look at the existence of equatorial circular orbits in the rKdS spacetimes.

\subsection{Existence of equatorial circular orbits} 
From relations (\ref{e16}) and (\ref{e17}) for specific energy and specific angular momentum of particles moving along equatorial circular orbits we conclude that such orbits can exist only in those regions in which the conditions
\be\label{e18}
\frac{1}{r^3}-y \geq 0,\quad 1-\frac{3}{r}\pm 2a\sqrt{\frac{1}{r^3}-y}>0
\ee
are fulfilled simultaneously.

The first inequality (\ref{e18}) determines the upper limit given by the \emph{static radius} $r_{\rm s}=y^{-1/3}$. Its existence is given by an interplay between gravitational attraction of the center and cosmic repulsion characterized by the cosmological constant $\Lambda>0$ \cite{Stu:1983:BULAI:}. As follows directly from Carter's equations (\ref{e6})-(\ref{e9}) and relations (\ref{e16}), (\ref{e17}), a~geodesic observer located at the static radius has only the time component of its 4-velocity non-zero: $u^{\mu}=(u^t,\,0,\,0,\,0)$; its position is, however, unstable against radial perturbations there. Location of the static radius in the rKdS spacetime is the same as in the standard \KdS or \SdS spacetimes \cite{Stu-Hle:1999:PHYSR4:,Stu-Sla:2004:PHYSR4:}, and is given by the condition
\be\label{e19}
y=y_{\rm s}(r)\equiv \frac{1}{r^3}.
\ee
Clearly, equatorial circular orbits exist only in the region $r\leq r_{\rm s}$. 

The second condition (\ref{e18}) is connected with limits given by photon circular orbits in the equatorial plane. Plus-family circular orbits exist only in the region where the rotational parameter $a>a_{\rm ph}^{+}(r;\,y)$, while the minus-family circular orbits exist only in the region in which the rotational parameter $a<a_{\rm ph}^{-}(r;\,y)$, where
\be\label{e20}
a_{\rm ph}^{\pm}(r;\,y)=\pm\frac{r(3-r)}{2\sqrt{r(1-yr^3)}}.
\ee 
Positions of plus/minus-family photon orbits are determined by equations $a=a_{\rm ph}^{\pm}(r;\,y)$, assuming $a\geq 0$. 

Clearly, for $y\leq 1/27$ the plus-family photon circular orbits are located in the region $0<r\leq 3$, while the minus-family ones can be found in the region $3<r<r_{\rm s}$. In black-hole backgrounds, there are two plus-family photon orbits, $r_{\rm ph1}^+$ located under the inner black-hole horizon and $r_{\rm ph2}^+$ located above the outer horizon, and one minus-family photon orbit $r_{\rm ph}^{-}>r_{\rm ph2}^+$ located above the outer horizon. In naked-singularity backgrounds, there is only one minus-family circular photon orbit $r_{\rm ph}^-$ and no plus-family ones. 

In rKdS spacetimes with the cosmological parameter $y>1/27$, no minus-family photon (as well as test-particle) circular orbits can exist; there are only plus-family photon and particle orbits in the region $0<r<r_{\rm s}$. In black-hole backgrounds, there are three plus-family photon orbits, $r_{\rm ph1}^{+}$ located under the inner horizon and $r_{\rm ph2}^{+}<r_{\rm ph3}^{+}$ located above the outer black-hole horizon. In naked-singularity backgrounds, there is only one plus-family photon orbit $r_{\rm ph}^+$. 

Regions of existence of test-particle equatorial circular orbits of a~given family in dependence of the cosmological parameter $y$ and character of rKdS spacetime (black hole (BH) or naked singularity (NS)) are presented in the Table \ref{t1}. Note that the limit on the existence of minus-family circular orbits given by the value of the cosmological parameter $y=1/27$ is the same as in the KdS spacetime. On the other hand, locations of photon orbits in rKdS spacetimes, given regardless of the family of orbits by the condition
\be\label{e21}
y=y_{\rm ph}(r;\,a)\equiv\frac{1}{r^3}-\left(\frac{r-3}{2ar}\right)^2,
\ee
differ from locations of photon orbits in KdS spacetimes with the same parameters $(a,\,y)$ \cite{Stu-Hle:2000:CLAQG:,Stu-Sla:2004:PHYSR4:,Sla-Stu:2020:EPJC:}.

\begin{table}%
\begin{tabular}{|c|c|c|c|c|} \hline
& \multicolumn{2}{c|}{$y\leq 1/27$} & \multicolumn{2}{c|}{$y>1/27$} \\ \hline
& BH & NS & BH & NS \\ \hline
& $0<r<r_{\rm ph1}^+$ & & $0<r<r_{\rm ph1}^+$ & \\
\raisebox{1.5ex}[0pt]{$+$} & $r_{\rm ph2}^{+}<r\leq r_{\rm s}$ & \raisebox{1.5ex}[0pt]{$0<r\leq r_{\rm s}$} & $r_{\rm ph2}^{+}<r<r_{\rm ph3}^+$ & \raisebox{1.5ex}[0pt]{$0<r<r_{\rm ph}^+$} \\ \hline
$-$ & $r_{\rm ph}^{-}<r\leq r_{\rm s}$ & $r_{\rm ph}^{-}<r\leq r_{\rm s}$ & no & no \\ \hline
\end{tabular}
\caption{Regions of existence of test-particle equatorial circular orbits of $+/-$ family in dependence of the cosmological parameter $y$ and character of rKdS spacetime (black hole (BH) or naked singularity (NS)). Limits are given by static radius $r_{\rm s}$ and/or circular photon orbit(s) $r_{\rm ph}$ of given family.}
\label{t1}
\end{table}

\subsection{Keplerian angular velocity and orientation of circular orbits}
Angular velocity of a~particle moving along equatorial circular orbit--so-called Keplerian angular velocity or Keplerian frequency--is given by the relation $\omega_{\rm K}=\d\phi/\d t=\dot{\phi}/\dot{t}$, where $\dot{t}$ and $\dot{\phi}$ are given by Carter's equations (\ref{e8}) and (\ref{e9}) evaluated for equatorial circular orbits with specific energies and specific angular momenta determined by relations (\ref{e16}) and (\ref{e17}). As a~result we obtain rather simple formula
\be\label{e22}
\omega_{\rm K\pm}=\frac{1}{a\pm r^{3/2}/\sqrt{1-yr^3}},
\ee
which is, quite unexpectably, the same as in the standard KdS spacetime \cite{Sla-Stu:2005:CLAQG:}. We can see that at the static radius $r_{\rm s}$ (in accord with its definition), $\omega_{\rm K}=0$. For physically relevant orbits at $r<r_{\rm s}$ (see Table \ref{t1}), $\omega_{\rm K+}>0$ while $\omega_{\rm K-}<0$. Therefore, particles orbiting on plus-family (minus-family) orbits are corotating (counterrotating) with the central object from the point of view of the static observer located at the static radius.

Other possibility, enabling to define, in fact locally, an orientation of orbits, uses the effect of frame dragging which takes place in all stationary and axially symmetric spacetimes, in which corresponding Killing vectors $\xi^{\mu}$ and $\eta^{\mu}$ are not orthogonal, i.e., $\xi_{\mu}\eta^{\mu}\neq 0$. The locally non-rotating frame (LNRF) is characterized by its angular velocity $\omega_{\rm LNRF}=-g_{t\phi}/g_{\phi\phi}$, which in the equatorial plane of the rKdS spacetime is given by the relation
\be\label{e23}
\omega_{\rm LNRF}=\frac{(2+yr^3)ar}{(r^2+a^2)^2-a^2\Delta}.
\ee
Clearly, $\omega_{\rm LNRF}>0$ even at the static radius. Particles, for which $\omega_{\rm K}>\omega_{\rm LNRF}$ perform, locally, prograde (or direct) motion, while particles on orbits with $\omega_{\rm K}<\omega_{\rm LNRF}$ perform retrograde motion with respect to the LNRF.\footnote{In papers \cite{Stu-Sla:2004:PHYSR4:,Sla-Stu:2005:CLAQG:,Sla-Stu:2020:EPJC:} discussing equatorial circular orbits in KdS or in even more general \KNdS backgrounds, prograde/retrograde orbits are called corotating/counterrotating instead.} In order to distinguish between prograde and retrograde orbits, it is sufficient to analyze sign of specific angular momentum of a~particle moving along given orbit. It is easy to show that $L=g_{\phi\phi}u^{t}(\omega_{\rm K}-\omega_{\rm LNRF})$, where $u^{t}=\d t/\d\tau$ is t-component of particle's 4-velocity $u^{\mu}$. Thus, the future-directed orbits, i.e., orbits with $u^{t}>0$, are prograde if $L>0$, and retrograde if $L<0$. From relations (\ref{e17}) we see that particles moving along minus-family orbits always perform retrograde motion with respect to the LNRF. On the other hand, particles on plus-family orbits exhibiting mostly prograde motion, can also move retrogradely. Typically, retrograde (with respect to the LNRF) but still corotating (with respect to the static observer) are plus-family orbits near the static radius. 

\subsection{Stability of equatorial circular orbits}
Equatorial circular orbits stable against radial perturbations (see the discussion on epicyclic motion in the next section) are determined by the condition $\d^2\tilde{R}/\d r^2\leq 0$, where equality holds for marginally stable orbits. Assuming $a\geq 0$, the marginally stable plus-family orbits are given by relations
\be\label{e24}
a=a_{\rm ms(1,2)}^{+}(r;\,y)\equiv\frac{\sqrt{r}}{3}\left\{4(1-yr^3)^{3/2}\pm\sqrt{\mathcal{D}}\right\},
\ee
while the minus-family ones are given by the relation
\be\label{e25}
a=a_{\rm ms}^{-}(r;\,y)\equiv\frac{\sqrt{r}}{3}\left\{-4(1-yr^3)^{3/2}+\sqrt{\mathcal{D}}\right\},
\ee
where
\be\label{e26}
\mathcal{D}=(1-4yr^3)[3r-2-yr^3(11-4yr^3)].
\ee

\begin{figure}
\includegraphics[width=1 \hsize]{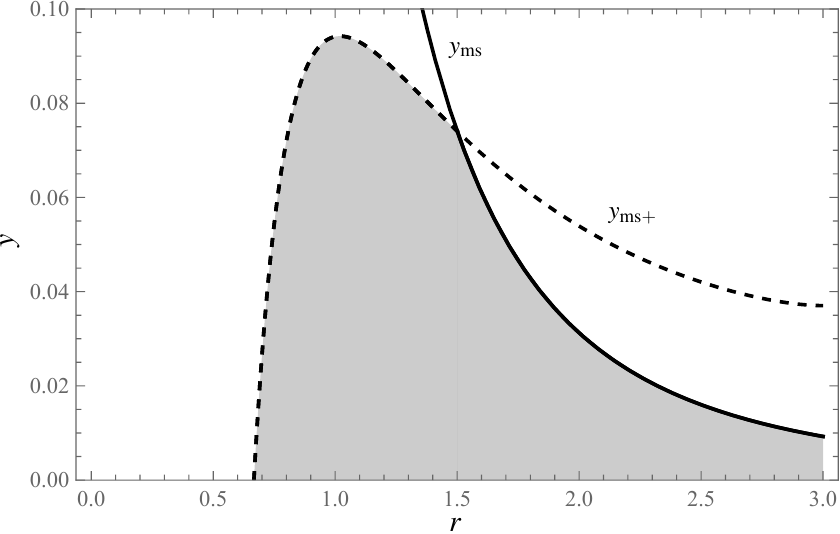}
\caption{Existence of stable equatorial circular orbits in dependence on the cosmological parameter $y$ of rKdS spacetimes.}
\label{f4}
\end{figure}

Reality conditions of both functions (\ref{e24}) and (\ref{e25}) determine regions of rKdS spacetimes where motion along stable equatorial circular orbits is possible. Their analysis reveals that stable circular orbits exist in those rKdS spacetimes, which cosmological parameter $y$ satisfies simultaneously following conditions
\bea
y &\leq& y_{\rm ms}(r)\equiv \frac{1}{4r^3}, \label{e27}\\
y &\leq& y_{\rm ms+}(r)\equiv\frac{1}{8r^3}[11-\sqrt{3(51-16r)}], \label{e28}
\eea
see Fig.~\ref{f4}. Maximum of the function $y_{\rm ms+}(r)$ determines maximal value of the cosmological parameter $y$ enabling existence of stable equatorial circular orbits in rKdS spacetimes. It corresponds to the upper limit for the existence of stable plus-family orbits as well: $y_{\rm c(ms+)}=12500/132651\doteq 0.09423$. Note that this value is a little bit higher than the corresponding limit for the existence of stable plus-family orbits in the KdS spacetime, which is $y_{\rm c(ms+)}^{\rm KdS}\doteq 0.06886$. On the other hand, stable minus-family orbits can exist only in rKdS spacetimes with the cosmological parameter $y<y_{\rm c(ms-)}=12/15^4$; this is the same limit as in the KdS spacetime and corresponds to the limit for the existence of stable circular orbits in the \SdS spacetime \cite{Stu-Sla:2004:PHYSR4:,Stu-Hle:1999:PHYSR4:}.

Motion along stable circular orbits of a given family is limited to the region $r_{\rm ms(i)}<r<r_{\rm ms(o)}$, where $r_{\rm ms(i)}$ ($r_{\rm ms(o)}$) correspond to the inner (outer) marginally stable orbits of a given family. In black-hole spacetimes allowing stable circular orbits, these orbits are located above the outer black-hole horizon only. Circular orbits at $r<r_{\rm ms(i)}$ and $r>r_{\rm ms(o)}$ are unstable ones. In rKdS spacetimes with both families of stable circular orbits, the region of stable minus-family orbits lie inside the region of stable plus-family orbits. Analysis of the function $a_{\rm ms(2)}^{+}(r;\,y)$, determining positions of marginally stable plus-family orbits in black-hole spacetimes (not only but also), reveals that the maximal value of the cosmological parameter $y$, allowing stable circular orbits around rKdS black holes, is $y_{\rm c(ms+)}^{\rm BH}=25/232\doteq 0.05787$.

Note that rising parts of $E_{\pm}(r)$ radial profiles as well as rising (descending) parts of $L_{+}(r)$ ($L_{-}(r)$) radial profiles correspond to stable circular orbits, and local extrema of both $E_{\pm}(r)$ and $L_{\pm}(r)$ determines positions and corresponding values of specific energies and specific angular momenta for marginally stable orbits, see Figs. \ref{f2}, \ref{f3}.

\section{Epicyclic motion}
In this section we shall consider small linear perturbations of particle's motion along equatorial circular orbits. We distinguish two independent perturbations--in radial and vertical directions, leading to the radial and vertical harmonic oscillations of a~particle around perturbed stable circular orbit, respectively.

\subsection{Radial epicyclic frequency}
We start from the equation of radial motion in the equatorial plane (\ref{e14}) where, now, we are interested in the rate of change of the radius $r$ in terms of the coordinate time $t$ instead of the proper time $\tau$. Clearly, we have
\be\label{e29}
\left(\frac{\d r}{\d t}\right)^2=\frac{1}{r^4 \dot{t}^2}\tilde{R}(r);
\ee
here $\dot{t}=\d t/\d\tau$. Subsequent time derivation gives the equation
\be\label{e30}
\frac{\d^2 r}{\d t^2}=\frac{\tilde{R}}{2}\frac{\p}{\p r}\left(\frac{1}{r^4 \dot{t}^2}\right)+\frac{\tilde{R}'}{2r^4 \dot{t}^2},
\ee
where $\tilde{R}'=\d\tilde{R}/\d r$.

Now we consider small radial displacement $\delta r$ around the mean equatorial circular orbit of the radius $r_0$, i.e., $r=r_0 + \delta r$. Since at the circular orbit $\tilde{R}(r_0)=\tilde{R}'(r_0)=0$, restricting to the linear terms in $\delta r$ the evolutionary equation for the radial displacement reads
\be\label{e31}
\frac{\d^2\delta r}{\d t^2}+\omega_{\rm R}^2\delta r=0,
\ee
where
\be\label{e32}
\omega_{\rm R}^2=-\left.\frac{\tilde{R}''}{2r^4 \dot{t}^2} \right|_{r_0}.
\ee
We see that linear perturbations of stable circular orbits (of both families), for which $\tilde{R}''<0$, leads to radial harmonic oscillations around the mean circular orbit with the radial epicyclic frequency given by the relation
\bea\label{e33}
\lefteqn{\omega_{\rm R\pm}=\frac{\omega_{\rm K\pm}}{\sqrt{1-yr^3}}\ \times} \\
& & \sqrt{1-\frac{6}{r}-yr^2(4r-15)\pm 8a\left(\frac{1-yr^3}{r}\right)^{3/2}-\frac{3a^2}{r^2}}, \nonumber
\eea
where $\omega_{\rm K}$ determined by relation (\ref{e22}) is the Keplerian orbital frequency for the given circular orbit of radius $r$ (in fact, $r=r_0$ in previous relation but for simplicity, we omit the subscript ``0'').

Due to conservation of angular momentum, the radial oscillations with angular frequency $\omega_{\rm R}$ leads, in the first order approximation, also to oscillations in azimuthal direction with the same frequency but different amplitude. Thus, the resulting motion of a particle can be viewed as a combination of elliptical motion along an epicycle and orbital motion of its center along the mean circular orbit, resembling ancient epicycle--deferent model of planetary motion used by Hipparchus and Ptolemy.  

\subsection{Vertical epicyclic frequency}
Equation (\ref{e7}) for particle's latitudinal motion, rewritten in terms of coordinate time $t$ instead of affine parameter $\lambda$, takes the form
\be\label{e34}
\left(\frac{\d\theta}{\d t}\right)^2=\frac{1}{\rho^4 \dot{t}^2}\tilde{W}(\theta),
\ee
where
\be\label{e35}
\tilde{W}(\theta)=\frac{\mathcal{K}}{m^2}-a^2\cos^2\theta-\frac{1}{\sin^2\theta}\left(aE\sin^2\theta-L\right)^2.
\ee
For the second order time derivative of coordinate $\theta$ we have the equation
\be\label{e36}
\frac{\d^2 \theta}{\d t^2}=\frac{\tilde{W}}{2}\frac{\p}{\p\theta}\left(\frac{1}{\rho^4 \dot{t}^2}\right)+\frac{\tilde{W}'}{2\rho^4 \dot{t}^2},
\ee 
where $\tilde{W}'=\d\tilde{W}/\d\theta$.

Now we consider small latitudinal perturbation $\delta\theta$ from the mean equatorial circular orbit, i.e., $\theta=\pi/2 + \delta\theta$. Since in the equatorial plane $\tilde{W}(\pi/2)=\tilde{W}'(\pi/2)=0$, restricting to the linear terms in $\delta\theta$ the evolutionary equation for the latitudinal perturbation reads
\be\label{e37}
\frac{\d^2\delta\theta}{\d t^2}+\omega_{\rm V}^2\delta\theta=0,
\ee
where
\be\label{e38}
\omega_{\rm V}^2=-\left.\frac{\tilde{W}''}{2\rho^4 \dot{t}^2} \right|_{\pi/2}.
\ee
It can be shown that $\tilde{W}''(\pi/2)=-2[a^2(1-E^2)+L^2]$ and that for any stable equatorial circular orbit, $E^{2}<1$, i.e., $\tilde{W}''(\pi/2)<0$. Thus, a~linearly perturbed particle performs vertical harmonic oscillations around the mean stable equatorial circular orbit with the vertical epicyclic frequency\footnote{The attribute `epicyclic' is generally used for every harmonic motion following from the first order perturbations of circular motion, especially in the study of stable accretion disc oscillations \cite{Now-Leh:1998:TheoryBlackHoleAccretionDisks:}.} given by the relation
\bea\label{e39}
\lefteqn{\omega_{\rm V\pm}=\frac{\omega_{\rm K\pm}}{\sqrt{1-yr^3}}\ \times} \\
& & \sqrt{1-yr^3\mp \frac{2a}{r^{3/2}}(2+yr^3)\sqrt{1-yr^3}+\frac{3a^2}{r^2}}. \nonumber
\eea

Note that the criteria of stability of equatorial circular orbits against radial and/or vertical perturbations are determined by the conditions $\omega_{\rm R}^2>0$ and/or $\omega_{\rm V}^2>0$. 

\begin{figure*}
\centering
\includegraphics[width=.95 \hsize]{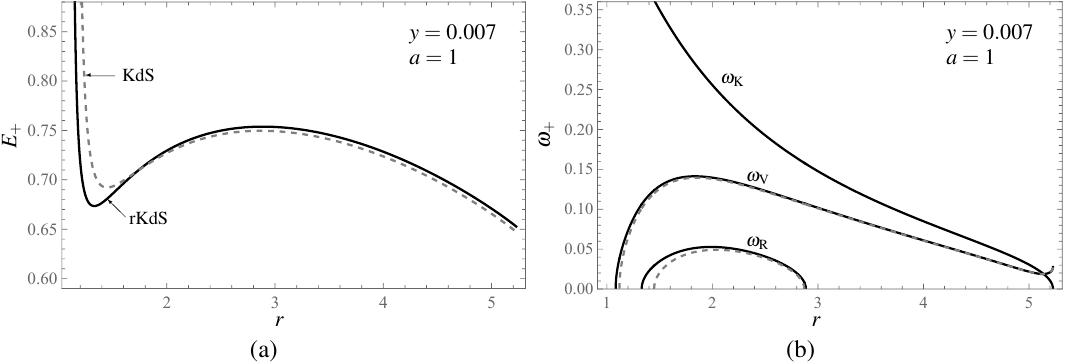}
\caption{(a) Radial profiles of specific energy for plus-family orbits in rKdS (solid) and KdS (dashed) black-hole spacetimes with the same spacetime parameters $y,\,a$. (b) Radial profiles of Keplerian ($\omega_{\rm K}$), radial epicyclic ($\omega_{\rm R}$) and vertical epicyclic ($\omega_{\rm V}$) frequencies for plus-family orbits in rKdS (solid) and KdS (dashed) black-hole spacetimes with the same spacetime parameters $y,\,a$.}
\label{f5}
\end{figure*}

\section{Conclusions}
Motion of massive test particles along equatorial circular orbits (geodesics) in the revisited KdS (rKdS) spacetime is studied for both black holes and naked singularities. The rKdS spacetime is a~new axially symmetric asymptotically de~Sitter solution of Einstein field equations. In contrast with the standard \KdS (KdS) solution, the rKdS metric in Boyer-Lindquist coordinates does not contain terms proportional to $ya^2$. Thus, the rKdS metric could be viewed as a limit of KdS metric for $ya^2\ll 1$. For illustration we present relations for specific energy and specific angular momentum of particles on equatorial circular orbits in standard KdS spacetime \cite{Stu-Sla:2004:PHYSR4:}:
\bea
E_{\pm} &=& \frac{\displaystyle 1-\frac{2}{r}-yr^2-\boxed{ya^2}\pm a\sqrt{\frac{1}{r^3}-y}}{\displaystyle\left[ 1-\frac{3}{r}-\boxed{ya^2}\pm 2a\sqrt{\frac{1}{r^3}-y}\right]^{1/2}}, \label{e40} \\
L_{\pm} &=& \frac{\displaystyle -\frac{2a}{r}-yar^2-\boxed{ya^3}\pm (r^2+a^2)\sqrt{\frac{1}{r^3}-y}}{\displaystyle\left[ 1-\frac{3}{r}-\boxed{ya^2}\pm 2a\sqrt{\frac{1}{r^3}-y}\right]^{1/2}}. \label{e41}
\eea
Although the rKdS metric and corresponding geodesic equations are simpler, the properties of equatorial circular orbits are, qualitatively, the same as in the KdS spacetime. Of course, there are some quantitative differences, for example in limits of existence and stability of circular orbits or in values of energy and angular momentum for given orbit, which, in principle, could be tested. Fig.~\ref{f5}(a) presents radial profiles of specific energy for plus-family orbits in rKdS and KdS black-hole spacetimes with the same spacetime parameters $y,\,a$. The main differences are in the location of the inner marginally stable orbit (given by locations of local minima in the profiles) and in the accretion efficiency $\eta$ given by the difference of specific energies of particles on the outer (given by location of the local maximum) and the inner marginally stable orbits, $\eta=E_{\rm ms(o)}-E_{\rm ms(i)}$. Clearly, in the rKdS spacetime, the accretion efficiency for corotating orbits is larger than in the standard KdS spacetime.

Motion along stable equatorial circular orbit, characterized by Keplerian orbital frequency $\omega_{\rm K}$, can be perturbed in both radial and vertical directions. Considering only small linear perturbations, the particle begin to oscillate in radial and vertical directions with radial and vertical epicyclic frequencies $\omega_{\rm R}$ and $\omega_{\rm V}$. The resulting motion is, thus, a~certain combination of three periodic motions with harmonic frequencies $\omega_{\rm K}$, $\omega_{\rm R}$ and $\omega_{\rm V}$, leading to the orbit which is still bound but, in general, not closed. However, for some orbits there should be resonances between these frequencies, characterized by ratios of small integers, which, for example, enable to investigate (quasi-)periodic phenomena in a~given background. In order to compare obtained results with their counterparts in the KdS spacetime, we also present expressions for epicyclic frequencies of orbiting particles in the standard KdS spacetime: 
\begin{widetext}
\bea
\omega_{\rm R\pm}&=&\frac{\omega_{\rm K\pm}}{\sqrt{1-yr^3}}\sqrt{1-\frac{6}{r}-yr^2(4r-15)\pm 8a\left(\frac{1-yr^3}{r}\right)^{3/2}-\frac{3a^2}{r^2}-\boxed{ya^2(1-4yr^3)}}, \label{e42}\\
\omega_{\rm V\pm}&=&\frac{\omega_{\rm K\pm}}{\sqrt{1-yr^3}}\sqrt{1-yr^3\mp \frac{2a}{r^{3/2}}(2+yr^3+\boxed{ya^2 r})\sqrt{1-yr^3}+\frac{3a^2}{r^2}+\boxed{\frac{ya^4}{r}}}. \label{e43}
\eea
\end{widetext}
Note that expressions for Keplerian frequency in both KdS and rKdS spacetimes are identical.
Fig.~\ref{f5}(b) shows radial profiles of three fundamental frequencies $\omega_{\rm K}$, $\omega_{\rm R}$ and $\omega_{\rm V}$ for particles on plus-family orbits in rKdS and KdS black-hole spacetimes with the same spacetime parameters $y,\,a$. We can see that in the case of corotating orbits with the same radial coordinate both epicyclic frequencies in the rKdS spacetime are higher than their counterparts in the KdS spacetime. On the other hand, the analysis of relations (\ref{e39}) and (\ref{e43}) reveals that in the case of counterrotating orbits the vertical epicyclic frequency in the rKdS spacetime is lower than the corresponding one in the KdS spacetime.

\begin{acknowledgments}
Support from the Research Centre for Theoretical Physics and Astrophysics, being a part of the Institute of Physics of the Silesian University in Opava, Czech Republic, is greatly acknowledged.
\end{acknowledgments}

\bibliographystyle{apsrev4-1}

%

\end{document}